\documentclass[10pt,journal,a4paper,twoside]{EWTEC}
\usepackage{cite}
\usepackage[pdftex]{graphicx}
\usepackage{textcomp}
\usepackage{amsmath}
\usepackage{threeparttable}
\usepackage[caption=false,font=footnotesize]{subfig}
\usepackage{url}

\begin{document}

\title{On the interaction of surface water waves and fully-submerged elastic plates}

\author{G. Polly,
        A. M\'{e}rigaud,
        R. Alhage,
        B. Thiria,
        and~R. Godoy-Diana
\thanks{In \emph{Proceedings of the 11th European Wave and Tidal Energy Conference} 5-9th Sept 2021, Plymouth, UK, Paper ID: 2221}
\thanks{Address: Physique et M\'ecanique des Milieux H\'et\'erog\`enes (PMMH), CNRS UMR 7636, ESPCI Paris--Universit\'e PSL, Sorbonne Universit\'e, Universit\'e de Paris, F-75005, Paris, France. e-mail addresses: gatien.polly@espci.fr, alexis.merigaud@espci.fr, rind.alhage@polytechnique.edu, benjamin.thiria@espci.fr, ramiro.godoy-diana@espci.fr}
} 

\markboth{}{POLLY \MakeLowercase{\textit{et al.}}: On the interaction of surface water waves and fully-submerged elastic plates}

\maketitle

\begin{abstract}
 The interaction between fully-submerged elastic plates and surface waves is of high interest to optimize wave energy converters such as the "Wave Carpet". This study focuses on the interaction between an elastic plate and waves from an experimental point of view. To do so, a submerged thin elastic plate is held at a fixed position in a wave field. The wave field is analyzed using a top view video recording that gives access to the water height field while the deformation of the elastic plate is recorded with a side view camera. Preliminary results presented in this paper show that the elastic plate transmits most of incoming wave energy while reflecting almost no energy. It is also shown that dissipation arises for wavelengths less than two times the length of the plate. Causes of dissipation are investigated emphasising that viscous dissipation could be responsible for the loss of energy.
\end{abstract}

\begin{EWTECkeywords}
Wave Carpet, Wave energy, Wave Energy Converter, Wave-structure interaction.
\end{EWTECkeywords}

%
\EWTECpeerreviewmaketitle

\section{Introduction}
\EWTECPARstart{H}ARVESTING ocean wave energy is of high interest in terms of applications and science. Yet, no wave energy converter (WEC) exploiting wave energy on a large scale and at relatively low cost has emerged, leading to the necessity of developing new designs. Apart from the creativity, necessary to elaborate new ideas, the development of new WEC designs needs numerical, analytical and experimental studies in order to optimize WEC properties.\\
Among new creative designs, the "Wave Carpet" is a promising WEC system composed of a flexible submerged plate attached to the seafloor. Pressure gradients induced by waves force the carpet to oscillate. The movement created actuate power take-off systems (PTO) producing electricity. The efficiency of such a system highly depends on the understanding of the fluid-structure interaction between waves and flexible plates.\\
This interaction has been largely documented from a numerical and analytical point of view. Prior to the idea of the Wave Carpet, the interaction between a flexible plate and a wave field has been studied to protect coastal region from waves \cite{Williams2003}, comparing submerged elastic plates to submerged rigid plates \cite{Cho2000}. First analytic solutions for non porous and porous elastic plates have been proposed by Cho and Kim \cite{Cho2000} \cite{Cho1998}, concluding that flexible plates could be efficient wave barriers. Mahmood-Ul-Hassan \textit{et al.} (2009) also solved the analytic problem for semi-infinite submerged thin plates \cite{Hassan2009}.
Alam (2012) first proposed to add PTO to flexible plates, and showed that the Wave Carpet could  be an efficient WEC by doing an analytical analysis of its behaviour \cite{Alam2012}. This observation gave rise to a growing interest for the study of the interaction between flexible plates and waves, leading to a significant number of numerical and analytical studies. Some of them directly focused on the Wave Carpet. For instance, Desmars \textit{et al.} (2018) investigated numerically the optimal PTO configuration \cite{Desmars2018} and Asaiean \textit{et al.} (2020) proposed another analytical description of the Wave Carpet and underlined that the mass of the plate is a key parameter in energy absorption \cite{Asaiean2020}. Several studies also focused on describing the interaction between a flexible plate and waves in different configurations or with different methods. Finally, Renzi (2016) explored analytically an alternative Wave Carpet design using piezoelectric material \cite{Renzi2016}.\\
The analytic problem of an elastic plate in a wave field has been solved by Mohapatra \textit{et al.} (2018) using both Green’s function technique and an eigenfunction expansion method \cite{Mohapatra2018}, and by Smith \textit{et al.} (2020) using the Wiener-Hopf technique \cite{Smith2020}. The effect of superposed elastic plates has also been modelled mathematically \cite{Mohapatra2014} \cite{Mohapatra2019} \cite{Behera2018}, as well as the presence of a wall placed near the plate \cite{Behera2015} and the effect of membrane porosity \cite{Gayathri2020} \cite{Guo2020} \cite{Mohapatra2020}.\\
However, very few experimental works have investigated the interaction between a flexible plate and waves. Cho and Kim (1998) completed their analytical and numerical study with measurements showing a good accuracy with the analytical development but some discrepancy could also be observed. The authors attributed this discrepancy to various effect that were not taken into account in the model, such as viscous damping \cite{Cho1998}. However, no further investigation on the causes of error has been made. Lehmann \textit{et al.} (2013) developed a model system of the Wave Carpet as a proof of concept underlining the efficiency of the Wave Carpet to absorb wave energy \cite{Lehmann2014}. Consequently, while there are many numerical and analytical studies, they are supported by few experimental work and the movement of the plate in the wave field has not be observed yet. The purpose of this study is to better understand the interaction between waves and a flexible plate from an experimental point of view.\\
In this study, a thin elastic plate is placed in monochromatic wave field and filmed from the top to obtain the deformation of the free surface using a Schlieren method \cite{Wildeman2018}, and from the side to observe the deformation of the membrane. This paper presents experimental and numerical methods that allow the determination of waves reflected and transmitted by the flexible plate. Preliminary results are also presented and suggest that energy dissipation is due to the interaction between waves and the flexible plate for wavelengths of the same order of magnitude than the length of the plate.

\section{Theoretical background}
\subsection{Linear wave theory}
Linear gravity wave theory consists in a linear description of waves in which the fluid is supposed to be incompressible and inviscid. The velocity field, \textbf{u}, is then described by Euler's equation:
\begin{equation}
    \rho_w(\partial_t\textbf{u}+(\textbf{u}.\nabla)\textbf{u})=\rho_w\textbf{g}-\nabla P
\end{equation}
where $\rho_w$ is the volumetric mass of water, $\textbf{g}$ the gravity and $P$ the pressure field. Under these assumptions the motion of the fluid is described by a potential, $\Phi$, defined as:
\begin{equation}
    \textbf{u}=\nabla\Phi
\end{equation}
where $\Phi$ is governed by the Laplace equation in the fluid domain:
\begin{equation}
    \Delta\Phi=0
\end{equation}
Considering a two-dimensional flow, the reference system (x,z) is chosen such that waves propagates in the $x$ direction. The $z$ direction is taken as the opposite direction of gravity and its origin is taken so that the free surface is located in $z=0$ as presented Fig.~\ref{Schemabassin}. Under the assumption of small amplitude waves, linearised boundary conditions are:
\begin{equation}
    \partial_z\Phi=0~~(z=-h)
\end{equation}
\begin{equation}
    \partial_t\eta=\partial_z\Phi~~(z=0)
\end{equation}
\begin{equation}
    \partial_t\Phi+g\eta=-\frac{P_0}{\rho_w}~~(z=0)
\end{equation}
Where $h$ stands for water depth, $\eta$ is the free surface elevation and $P_0$ is the pressure in $z=0$.
Considering a monochromatic wave without damping, the free surface elevation writes:
\begin{equation}
    \eta=\eta_0\mathrm{e}^{-\mathrm{i}(2\pi f t-kx)}
\end{equation}
where $\eta_0$ is the amplitude of the wave, $f$ its frequency and $k$ its wavenumber. $f$ and $k$ are related by the dispersion relation:
\begin{equation}
    (2\pi f)^2=gk\tanh(kh)
\end{equation}
Finally, the analytic expression of the potential $\Phi$ is:
\begin{equation}
    \Phi=-\frac{ig\eta_0}{\omega}\frac{\cosh(k(z+h))}{\cosh(kh)}\mathrm{e}^{-\mathrm{i}(\omega t-kx)}
\end{equation}

\subsection{Interaction between waves and an obstacle\label{Section theory2}}

\begin{figure}[!t]
\centering
\includegraphics[width=86mm]{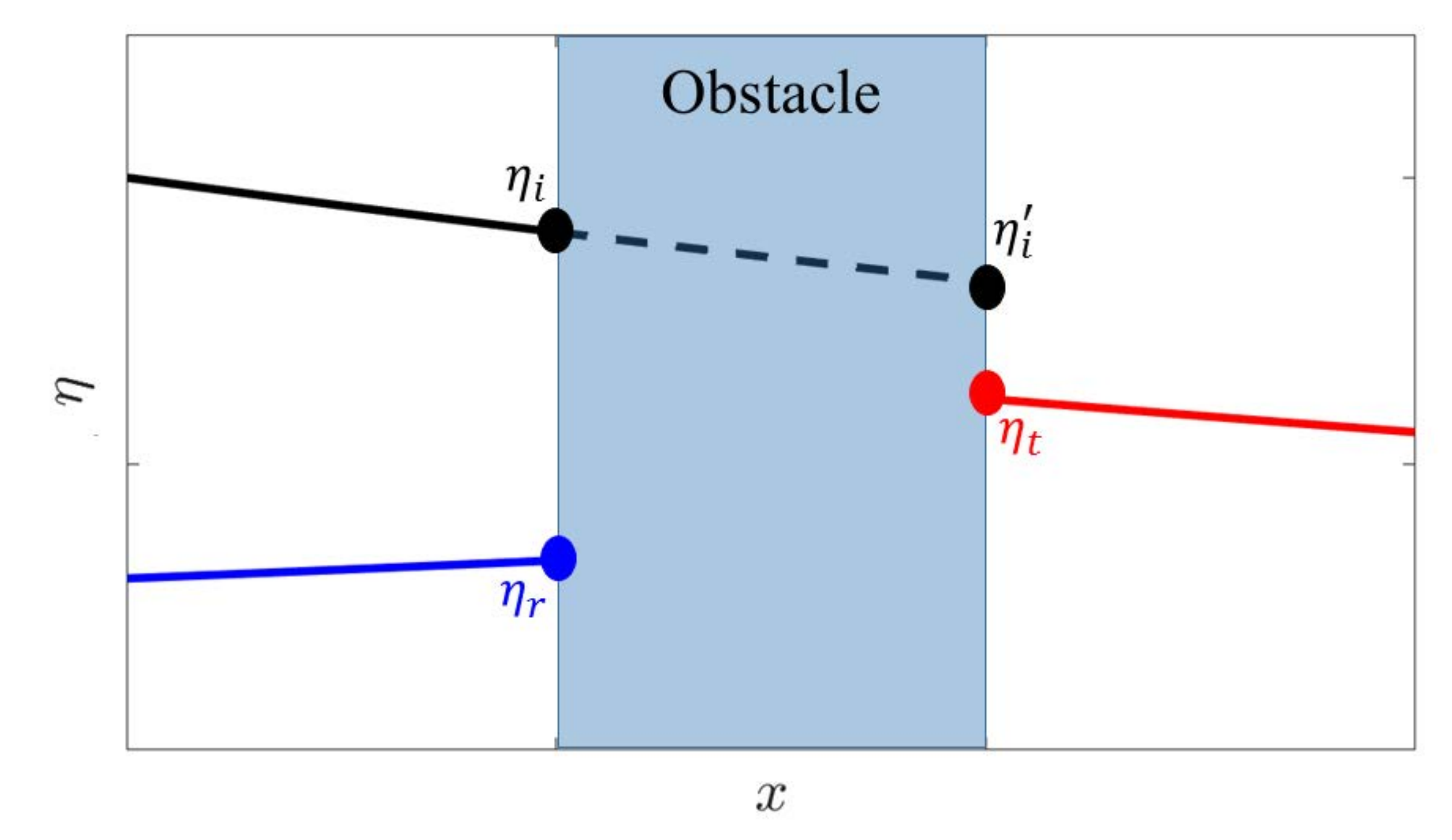}
\caption{Scheme picturing waves amplitude before and after the obstacle. Amplitude is pictured decreasing to represent dissipation that takes place in a wave tank.}
\label{Schema_amplitudes}
\end{figure}

Wave-structure interaction is a complex problem because of the strong coupling between the wave field and the object.\\
On the one hand, waves exert a force on the object. This force, $F_p$, can be expressed as the sum of pressure's efforts on the submerged part of the obstacle:
\begin{equation}
    F_p=\iint_SpdS
\end{equation}
where $S$ the submerged surface of the object and $p$ the dynamic pressure defined as:
\begin{equation}
    p=-\rho_w\partial_t\Phi
\end{equation}
So that:
\begin{equation}
    F_p=-\rho_w\iint_S\partial_t\Phi dS
\end{equation}
Where $\Phi$ is the sum  of the incident, diffracted and radiated potentials.
This force, generated by waves, leads to motion of the object, creation of added mass and radiation damping.
\\ On the other hand, the presence of the object influences the wave field leading to wave reflection and transmission by the obstacle. Before the object, the free surface elevation can be written as the sum of two waves, one propagating forward and one propagating backward:
\begin{equation}
    \eta=\eta_i\mathrm{e}^{-\mathrm{i}(\omega t-kx)}+\eta_r\mathrm{e}^{-\mathrm{i}(\omega t+kx)}
\end{equation}
where $\eta_i$ is the amplitude of the incident wave propagating forward and $\eta_r$ the amplitude of the reflected wave propagating backward. These two amplitudes are used to obtain the coefficient of reflection in amplitude, $K_r$, defined as $\eta_r/\eta_i$. After the object, only the transmitted wave propagates and the free surface elevation writes:
\begin{equation}
    \eta=\eta_t\mathrm{e}^{-\mathrm{i}(\omega t-kx)}
\end{equation}
where $\eta_t$ is the amplitude of the transmitted wave. The coefficient of transmission in amplitude, $K_t$, is defined as  $\eta_t/\eta_i'$  where $\eta_i'$ is the amplitude of the incident wave after the obstacle as defined in Fig.~\ref{Schema_amplitudes}. More precisely, $\eta_i'$, corresponds to the amplitude that the incident wave would have in the "downwaves" zone in the absence of the object. The measurement of $K_r$ and $K_t$ allows to characterize the interaction between the object and the wave field. Since wave energy is proportional to the square of amplitude, the coefficients of reflection and transmission in energy are defined as $K_r^2$ and $K_t^2$. If no dissipation takes place, the totality of incoming energy is split into reflection and transmission, so that $K_r^2+K_t^2=1$.\\
However, in general the energy of the incoming wave can also be dissipated due to two main processes:
\begin{itemize}
    \item internal dissipation: dissipation in the object which can be enhanced by using PTO and corresponds to the energy that could be harvested by the WEC for power production purposes. Considering a fully submerged thin elastic plate of length, $L$, fixed at two points, the energy dissipated by internal dissipation over one wave period, $E_a$, can be defined as:
    \begin{equation}
    E_a=2\Gamma\rho_peb\int_{0}^{T}\int_{0}^{L}(\partial_t\xi)^2\mathrm{d} x \mathrm{d} t
    \end{equation}
    where $\Gamma$ is the internal damping coefficient, $\rho_p$ is the density of the plate, $e$ its thickness, $b$ its width, $L$ its length, $T$ the wave period and $\xi$ the plate deflection profile.
    
    \item drag dissipation is the dissipation that arises in the fluid due to the creation of vortices. As described in Pi\~neirua \textit{et al.} (2017) for a fully submerged thin elastic plate of length, $L$, fixed at two points the power by unit of surface lost by drag, $P_s$, is expressed as \cite{Pineirua2017}:
    \begin{equation}
    P_s=-\frac{1}{2}C_D\rho_w|U_n|U_n^2
    \end{equation}
    where $C_D$ is the drag coefficient and $U_n$ the relative speed between the fluid and the plate normal to the plate. The energy lost by drag during one wave period, $E_d$, is obtained by integrating this expression over the surface of the plate and one period of wave:
    \begin{equation}
         E_d=-\frac{1}{2}C_D\rho_wb\int_0^T\int_0^L|U_n|U_n^2 \mathrm{d} t\mathrm{d} x
    \end{equation}
    where $b$ and $L$ are respectively the width and the length of the plate and $T$ is the wave period. Considering that the movement of the plate is vertical and of small amplitude, an expression of the relative velocity between the fluid and the object can be obtained:
    \begin{equation}
        U_n=\partial_t\xi-u_z-u_x\partial_x\xi
    \end{equation}
    where $u_z$ and $u_x$ are respectively the $z$ and the $x$ components of the velocity field.
\end{itemize}

The subsequent sections focus on the experimental determination of the energy balance between reflection, transmission and dissipation for a fully submerged elastic plate.

\section{Methods}
\subsection{Experimental setup}
\begin{figure}[!h]
\centering
\includegraphics[height=.21\textheight]{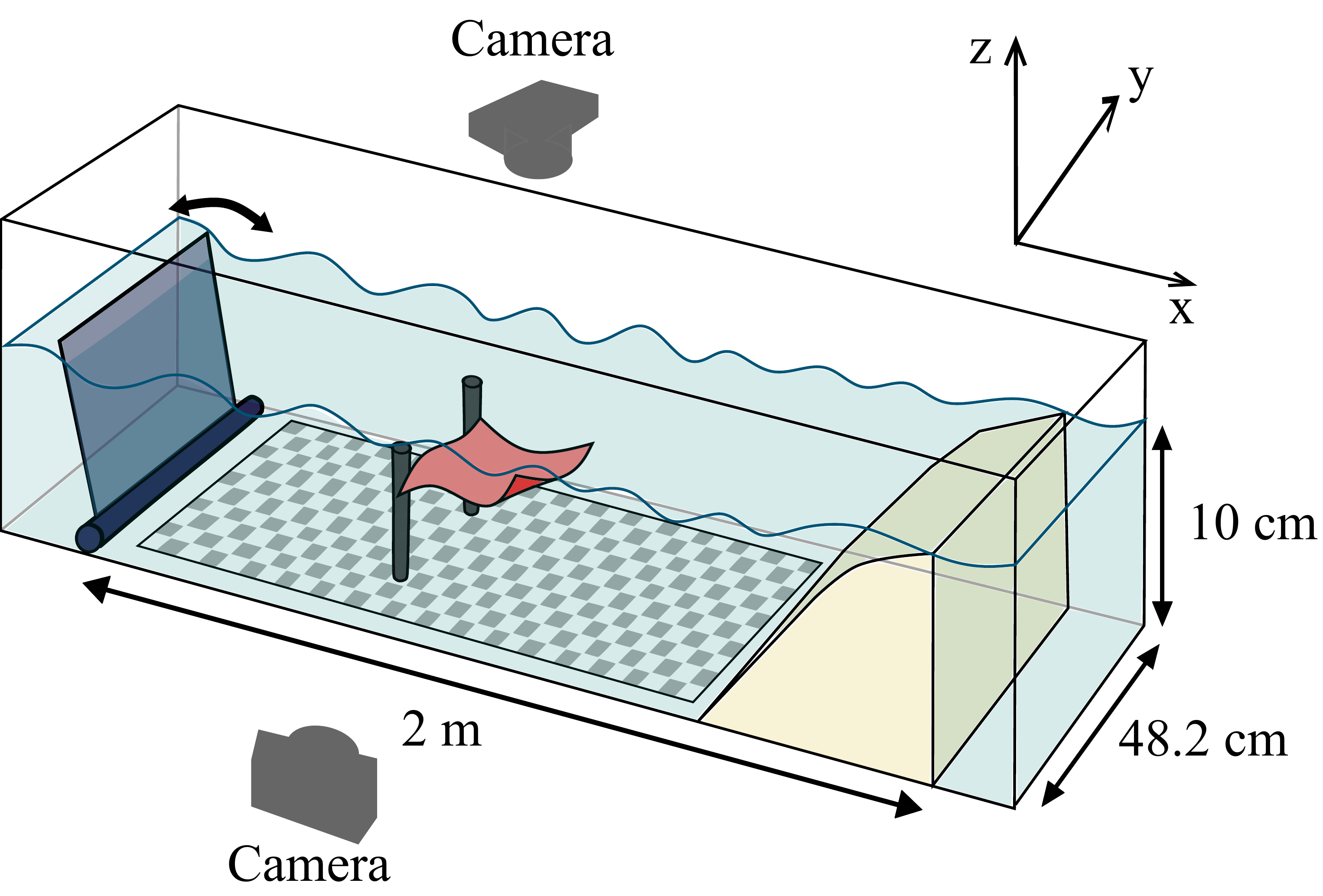}
\caption{Schematic diagram of the water tank. From left to right the wavemaker, the thin elastic plate and the beach are represented. The checkerboard pattern on the bottom is used to reconstruct the water height map using the top view camera (see text). The side view camera is used to track the deformation of the plate. \label{Schemabassin}}
\end{figure}
The experiments are performed in a wave tank of dimensions 2.5 x 0.48 x 0.2 m, where a flexible plate undulates in response to the forcing induced by the waves (see Fig.~\ref{Schemabassin}).
A flap-type wavemaker driven by a DC motor and a crank rod junction is used to generate waves. Its actuation is controlled using Matlab\footnote{www.mathworks.com} and an Arduino\footnote{www.arduino.cc} Mega  board. The waves produced can have frequencies from 1 to 5 Hz and amplitudes from 2 to 5 mm depending on the frequency. A beach is placed 2 m away from the wavemaker in order to reduce wave reflection.The beach has been designed and 3D printed using the study by Ouellet and Datta (1986) \cite{Ouellet1986},  which emphasized the efficiency of parabolic beaches pierced with holes. Water depth must be equal to the height of the beach in order to optimize wave absorption by allowing wave breaking. The water depth is 10 cm in all experiments in order to satisfy this condition.\\
The plate is connected to two points to support poles fixed to the bottom of the tank using LEGO\textregistered~plates located 1 m away from the beach and the wavemaker. The poles maintaining the membrane are elongated rectangular parallelepiped that do not disturb the wave field (see Appendix A). They are pierced in order to fix the thin plate at different depths. The plate width is 20 cm, 
\begin{figure}[!h]
\centering
\includegraphics[width=86mm]{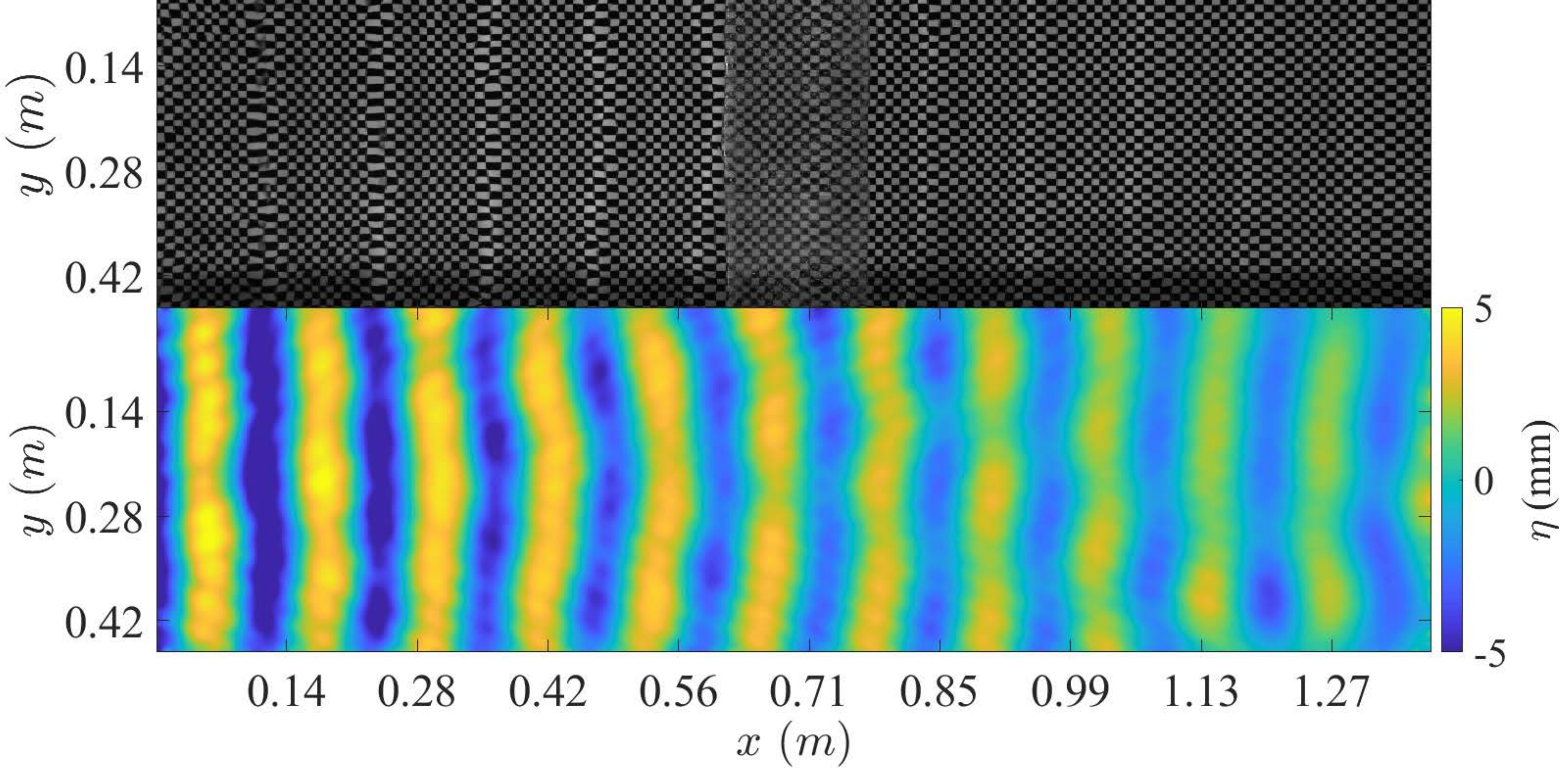}
\caption{Top: picture taken at the top of tank without membrane while waves of 3.7 Hz are generated. The grey surface at the center corresponds to LEGO plates used to fix the poles maintaining the membrane. Bottom: map of water elevation obtained using the Schlieren method \cite{Wildeman2018}. Yellow areas corresponds to the free surface height, $\eta$. For both images, x corresponds to wave’s direction of propagation and y to the width of the tank.}
\label{snapshoteta}
\end{figure}
its length is 12 cm so that waves produced have similar wavelength. 
Its thickness is taken equal to 57~$\mu$m and is small compared to every other dimension of the setup so that its effect is negligible. It is made of polypropylene to limit the difference of density with water.\\
Two Basler\footnote{www.baslerweb.com} acA200 cameras are used to record from the side and from the top in order to track, simultaneously, the movement of the membrane and the wave field. Both acquisition rates are fixed at 50 frames per second. A checkerboard made of 1 x 1cm squares is placed under the tank in order to measure the surface elevation using the Schlieren method as discussed in Wildeman (2018) and detailed in Section~\ref{Section Method 2} \cite{Wildeman2018}.

\subsection{Detection of waves and numerical treatment \label{Section Method 2}}

\begin{figure}[!t]
\centering
\includegraphics[width=86mm]{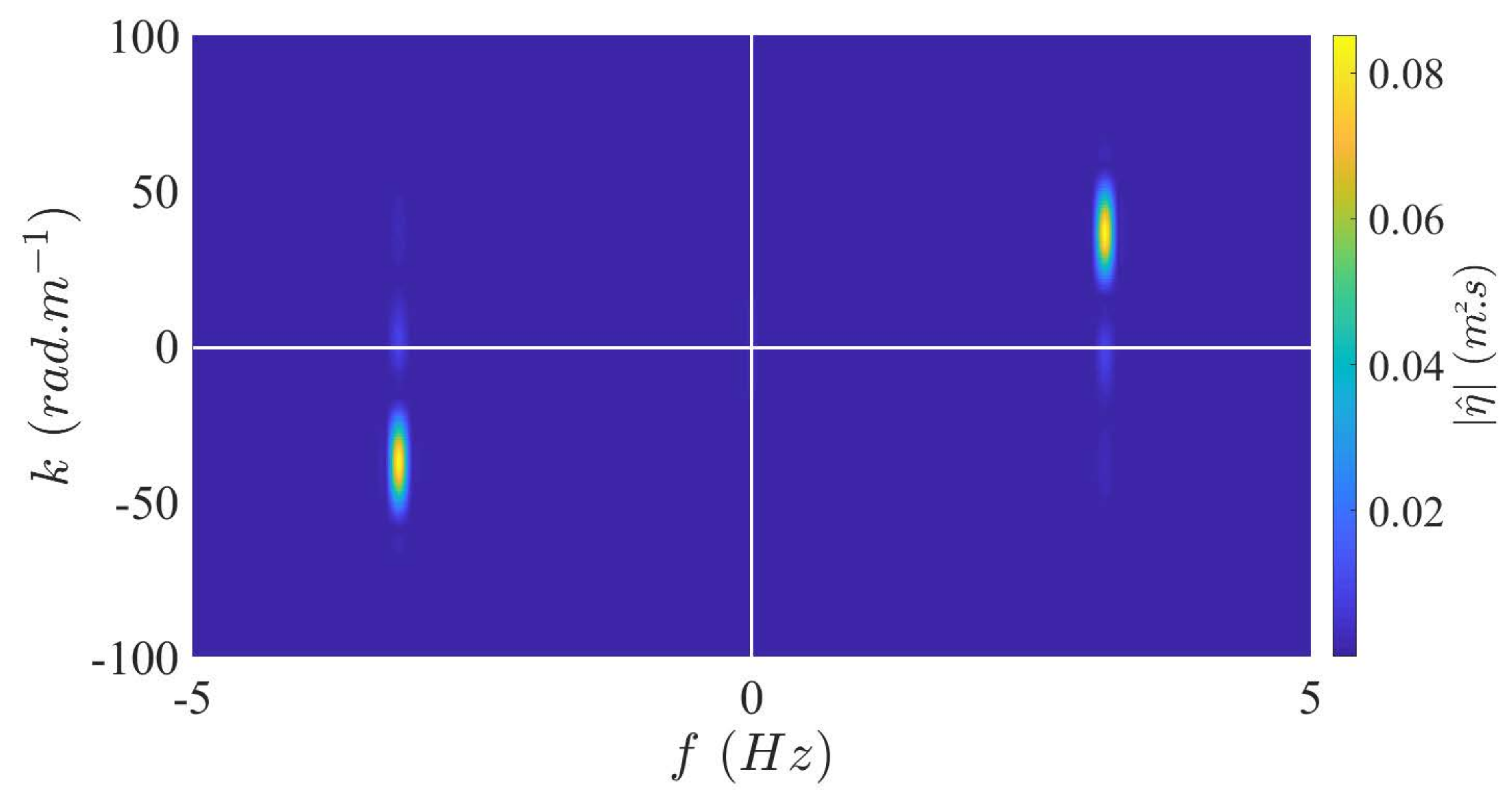}
\caption{Phase diagram of the upstream area obtained by performing a Fourier transform in space and in time for 3.5 Hz waves. Most of the energy is located in areas corresponding to wave propagating upward. Energy due to reflection by the flexible plate can also be observed for regions in which the wavenumber $k$ and the frequency $f$ don't have the same sign.}
\label{kw_space}
\end{figure}
The free surface elevation profile across the tank is determined using Schlieren method on each image of the movie \cite{Wildeman2018}. Knowing water depth at rest, the deformation of the checkerboard allows the calculation of the free surface deformation at any time as illustrated Fig.~\ref{snapshoteta}.\\
To characterize the wave-plate interaction the coefficients of transmission and reflection $K_t$ and $K_r$, as defined section~\ref{Section theory2}, are measured. To do so the free surface height, $\eta$, is averaged over the width of the tank and separated in two components corresponding to a waves propagating forward and backward in both "upwaves" and "downwaves" zones. This separation is usually performed using the method proposed by Mansard and Funke (1980) consisting in measuring simultaneously the surface height at three different locations in the tank and using a least square method \cite{Mansard1980}. However, in this study a method based on Fourier transform is used. Thanks to the Schlieren method the height of the free surface is known in all the tank at any time allowing to separate incident and reflected waves by using a filter in Fourier space. More precisely, a 2D Fourier transform in space and in time is performed on the matrix containing the averaged free surface height. It allows to obtain the distribution of energy in the phase space as illustrated Fig.~\ref{kw_space}.
\begin{figure}[!h]
\centering
\includegraphics[width=86mm]{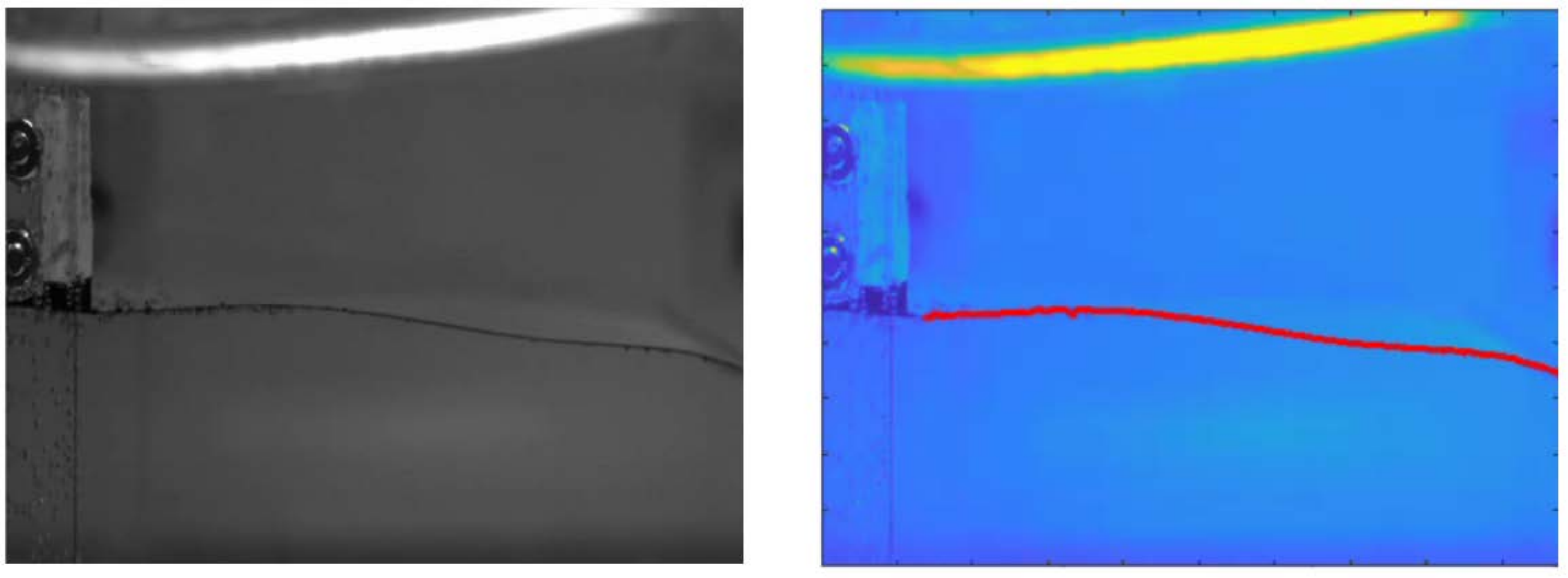}

\caption{Left: Side view of the thin plate in a wave field. The bright layer on the picture is due to the free surface. Right: Side view of the plate after treatment. The red line corresponds to the fitted edge of the plate.}
\label{FitMembrane8img}
\end{figure} The quadrants of this graph in which $k$ and $f$ have identical (respectively opposite) signs corresponds to waves propagating forward (respectively backward). To improve the accuracy of the 2D Fourier transform, a zero padding is used such that the length of the signal treated reaches ten times the length of the initial signal. As illustrated in Fig.~\ref{Schema_amplitudes}, using Fourier analysis allows to define the complex amplitude of the incident wave before and after the elastic plate, respectively $\eta_i$ and $\eta_i'$, the complex amplitude of the reflected wave, $\eta_r$, and the complex amplitude of the transmitted wave $\eta_t$.

\subsection{Elastic plate detection}

The elastic plate is recorded from the side. The precise movement of the membrane is extracted using a detection algorithm that consists of finding the darkest squares in a given area. As illustrated in Fig.~\ref{FitMembrane8img}, it allows to extract the plate deflection profile with satisfactory accuracy.

\section{Results}
\subsection{Reflexion and transmission coefficients in energy\label{Section: ResultsA}}

The influence of the thin elastic plate on the wave field can be characterized by measuring the reflection and transmission coefficients in energy for different wavelengths. To that end, multiple experiments are performed fixing membrane extremities at a 2 cm depth and varying the wavelength from 0.12 to 0.45 m. In each experiment the water is initially still and a monochromatic wave field is imposed during 1 minute and  30 seconds.
Movies from both the top and the side of the tank are then recorded simultaneously at 50 frames per second during the last 30 s so that a steady state has been reached. Complex amplitudes of incident, transmitted and reflected waves are then calculated following the procedure described in Section~\ref{Section Method 2} in order to obtain reflection and transmission coefficients $K_r^2$ and $K_t^2$. For every wavelength, the experiment is repeated five times.\\
\begin{figure}[!h]
\centering
\includegraphics[width=86mm]{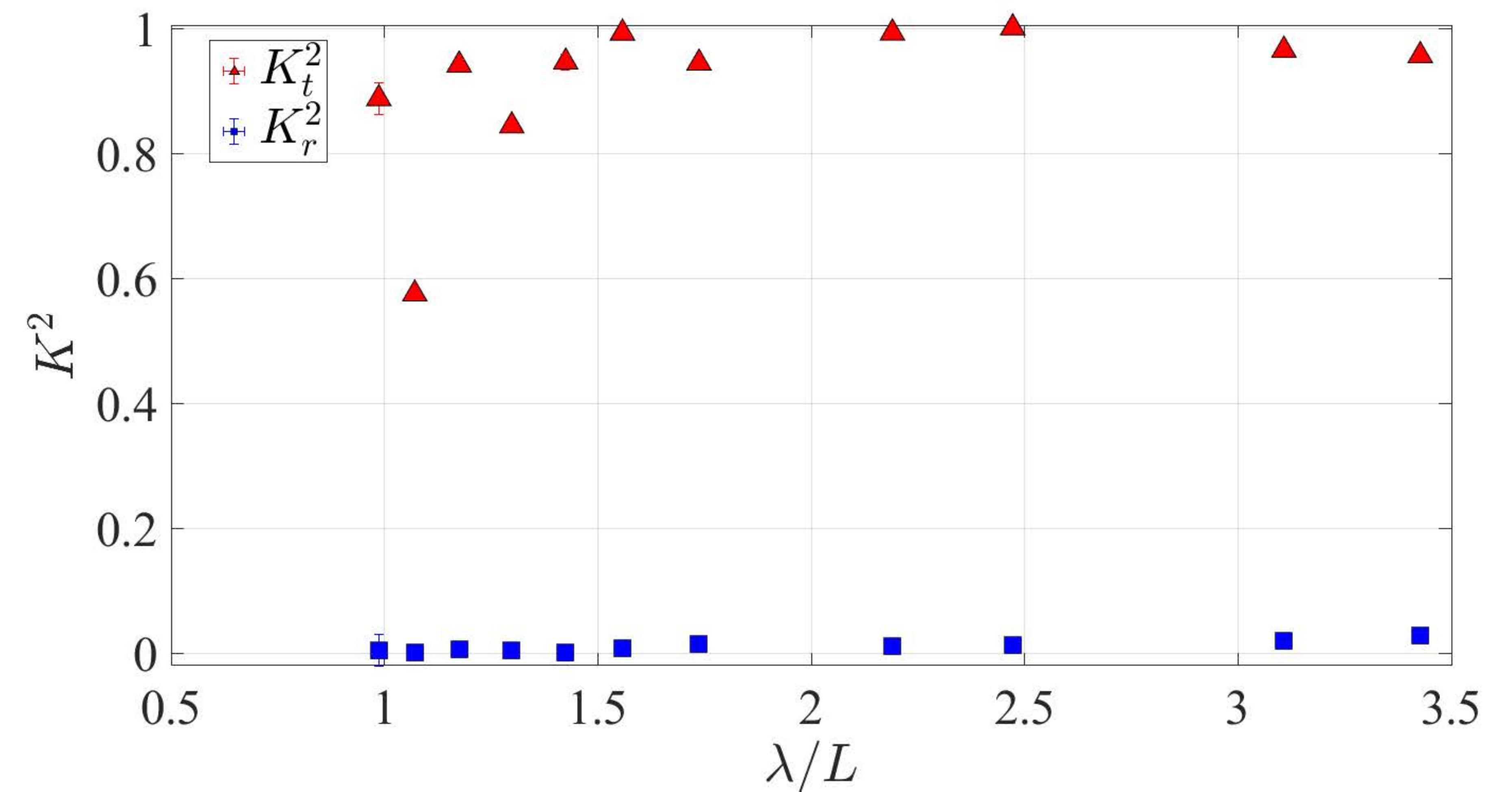}

\caption{Coefficients of reflection (blue squares) and transmission (red triangles) in energy as a function of wavelength, $\lambda$ normalized by the length of the plate, $L$. Most of the energy is transmitted by the elastic plate but for wavelengths almost equal to $L$ dissipation can be observed.}
\label{Coeff_nrj_times2}
\end{figure}
The results are shown Fig.~\ref{Coeff_nrj_times2} presenting mean value of the reflection and transmission coefficients in energy as a function of the reduced wavelength, $\lambda/L$, where $L$ is the length of the plate. Errorbars correspond to the standard deviation.

These results show that for wavelength longer than the plate length, the vast majority of incoming wave energy is transmitted, since the reflection coefficient is almost equal to zero for every wavelength. More importantly this graph shows two regimes. For wavelengths twice superior to the length of the plate, $L$, the transmission is total, and the presence of the plate leads to little energy dissipation (around 5\% of total incoming energy). However, for wavelengths shorter than twice the length of the elastic plate, transmission decreases significantly. Since no increase in reflection is observed, the missing energy can be attributed to an increase in dissipation.

\subsection{Dissipation and elastic plate dynamics in the wave field}

\begin{figure}[!h]
\centering
\includegraphics[width=86mm]{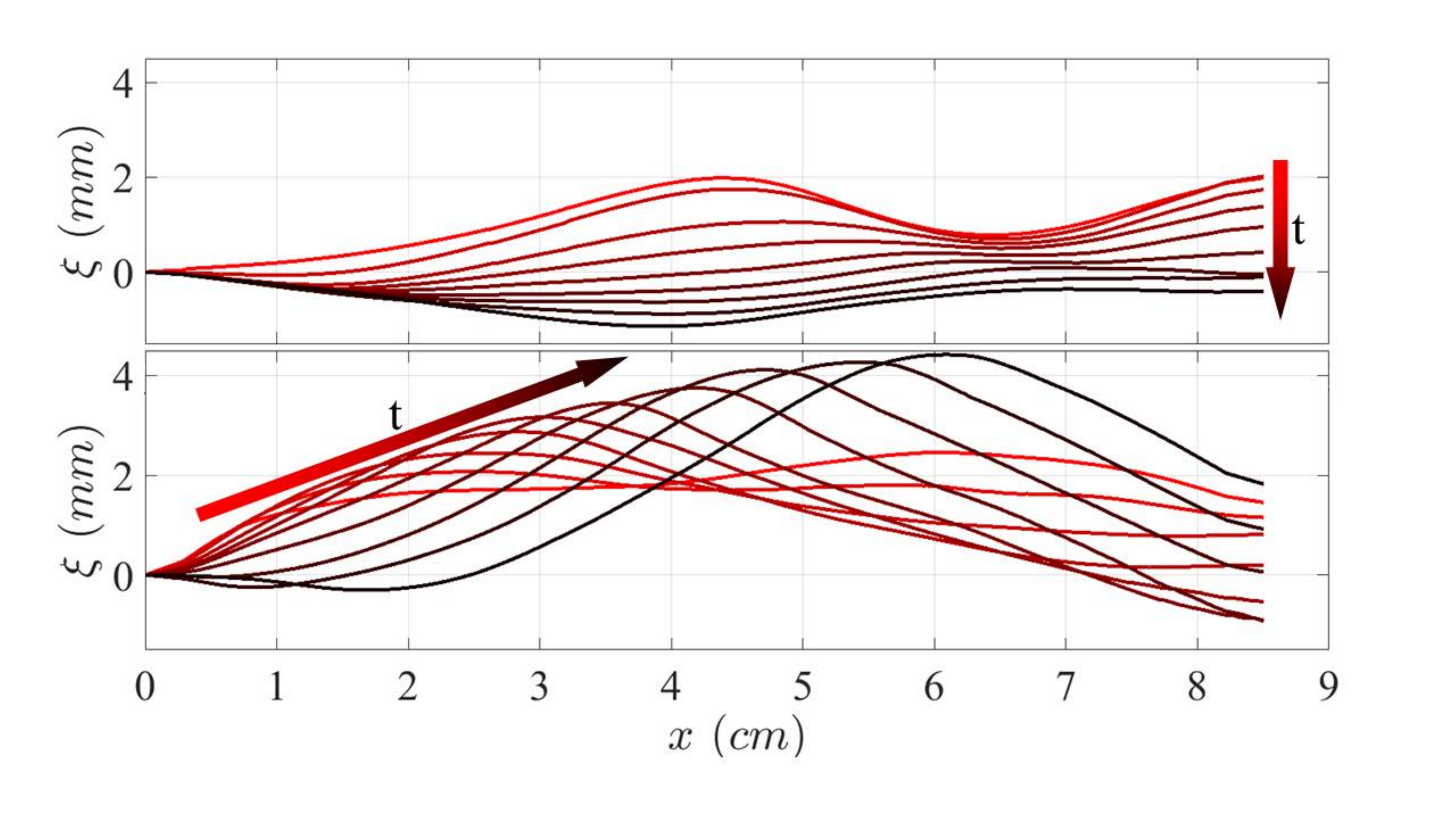}

\caption{\textit{Top:}  Elevation of the plate, $\xi$, as a function of the position at different times separted from 0.02 s going from red to black for $\lambda/L$=3.4. The curves exhibit nodes and anti-nodes characteristic of stationnary waves. \textit{Bottom:} Elevation of the plate, $\xi$, as a function of the position at different times separted from 0.02 s going from red to black for $\lambda/L$=1.07. It shows the apparition of a propagative waves in the elastic plate.}
\label{Forme_membrane}
\end{figure}

The plate deflection profile dynamics are now examined, to refine further the characterization of the two regimes identifiable in Section~\ref{Section: ResultsA}. A closer look at the plate shows that its movement is characteristic of the presence of strong damping for $\lambda$ close to $L$. More specifically as illustrated in Fig.~\ref{Forme_membrane}, progressive waves are propagating in the plate for $\lambda/L$=1.07 whereas for $\lambda/L$=3.4 stationary waves are observed. Since Ramananarivo \textit{et al.} (2014) highlighted that dissipative phenomena can lead to the apparition of propagative waves instead of stationary waves \cite{Ramananarivo2014}, the observation of Fig.~\ref{Forme_membrane} corroborates the findings of Fig.~\ref{Coeff_nrj_times2}, \textit{i.e.} that for $\lambda$ close to $L$ the thin elastic plate dissipate wave energy.

\subsection{Investigating the causes of dissipation: internal damping and drag\label{Section: ResultsC}}
At first sight three mechanisms can account for the loss of energy observed, internal and drag dissipation.
\subsubsection{Internal damping}
First of all, if the thin plate shows internal dissipation, some energy will be lost because of its oscillation. A free oscillation test performed in the air allows to define an internal damping coefficient, $\Gamma$, of 0.75 s$^{-1}$ (see Appendix B) with the assumption that damping in air is only due to internal dissipation. Using the expression of $E_a$, the energy dissipated by internal dissipation over one wave period, given Section~\ref{Section theory2} it can be computed as well as the energy of the incoming wave, $E_w$:
\begin{equation}
    E_w=\frac{1}{2}\rho_wgW\lambda|\eta_i|^2
\end{equation} 
where $\rho_w$ is the density of water, $g$ the acceleration of gravity, $W$ the width of the tank, $\lambda$ wave's wavelength and as defined in Section~\ref{Section theory2} $\eta_i$ stands for the complex amplitude of the incident wave. For most wavelengths the energy supposedly dissipated by the elastic plate has an order of magnitude of $10^{-7}$ J so that $E_a/E_w$ is almost equal to $10^{-4}$ and is very small compared to the observed dissipation. Consequently, internal damping in the plate cannot be responsible for the dissipation observed.

\subsubsection{Drag induced dissipation}
\begin{figure}[!h]
\centering
\includegraphics[width=86mm]{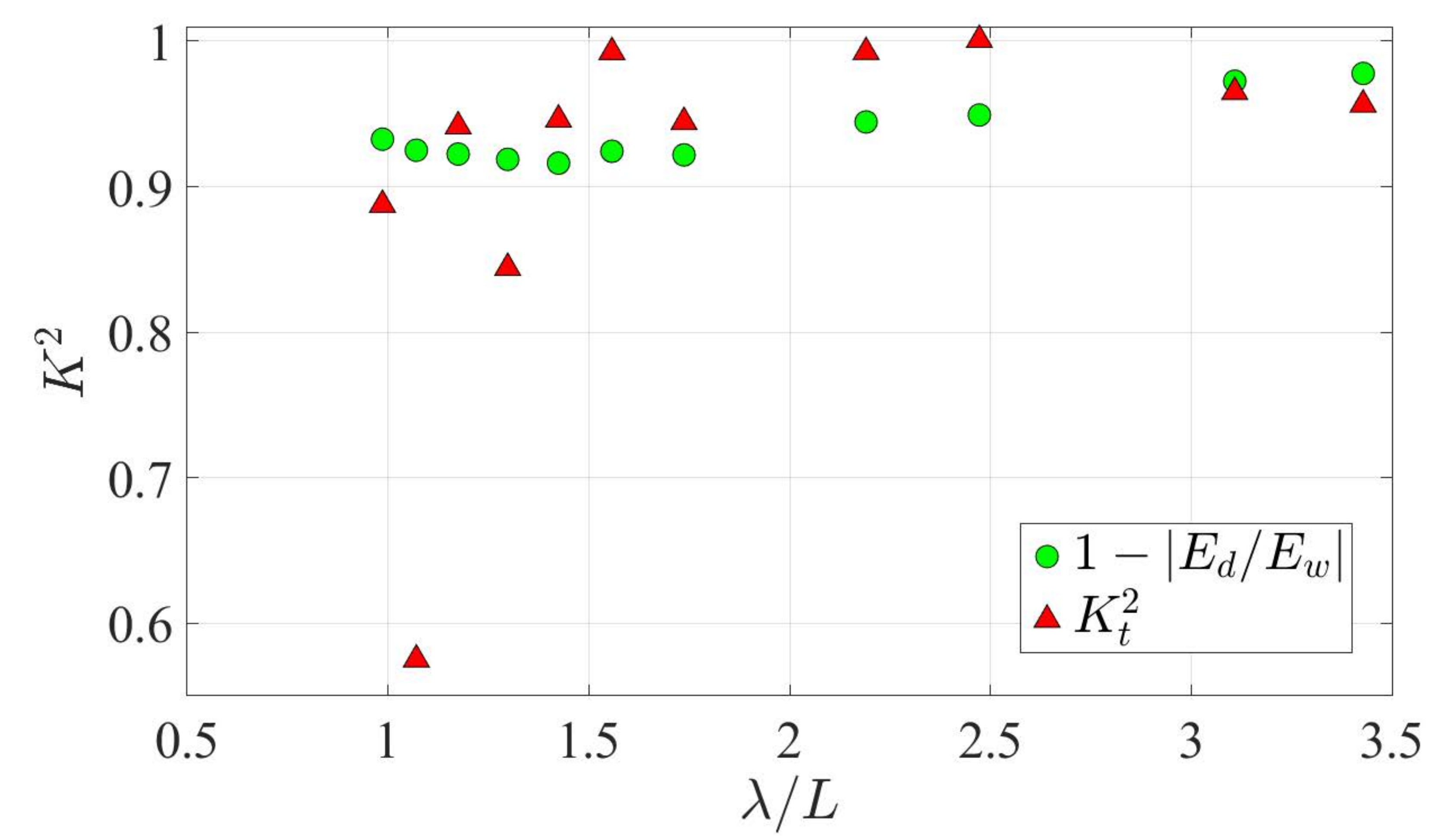}
\caption{Proportion of incoming wave energy dissipated according to the simple model developed, $1-|E_d/E_w|$, (green dots) and the coefficient of transmission in energy, $K_t^2$ as a function of the wavelength normalized by the length of the plate. Estimated dissipation has the same order of magnitude and is also more important for short wavelengths as observed experimentally.}
\label{dissipation_drag}
\end{figure}

Dissipation can also be due to the energy lost because of the creation of drag.
In order to obtain an order of magnitude of the energy dissipated because of the drag, $E_d$, over one wave period the assumption that $u_z$ and $u_x$ the $x$ and $z$ components of the velocity field are given by the linear theory is made and $\xi$ is assumed to have the following form: 
\begin{equation}
    \xi(x,t)=\xi_0\frac{\sinh(k_mx)}{\sinh(k_mL)}\mathrm{e}^{-i(2\pi f t-k_m x)}
\end{equation}
where $\xi_0$ is the elastic plate oscillation's amplitude for $x=L$ and $k_m$ is the wavenumber of the wave propagating in the membrane. $k_m$ is determined using the dispersion relation:
\begin{equation}
    k_m^2=2\pi f\sqrt{\frac{\mu}{EI}}
\end{equation}
where $E$ and $I$ are respectively the Young's modulus and the second moment of inertia of the plate. $\mu$ stands for the sum of the volumetric mass density of the plate and the added mass due to the motion of the plate. The added mass is determined using the expression given by Pi\~neirua \textit{et al.} (2015) \cite{Pineirua2015}.\\
The energy dissipated because of drag, $E_d$, can then be computed using trapezoidal method to calculate the integrals and the expression given Section~\ref{Section theory2}. The proportion of energy of the incident wave that is dissipated is equal to $1-|E_d/E_w|$. It is presented in Fig.~\ref{dissipation_drag} for wavelengths produced in the tank as well as $K_t^2$, the coefficient of transmission in energy, measured experimentally. Both quantities are of the same order of magnitude. Also, this simple model pictures an extremum in the dissipation for $\lambda\approx1.5 L$. 
This is particularly interesting since a similar trend is also observed experimentally in Fig.~\ref{Coeff_nrj_times2} but with larger magnitude.
 The present theoretical model exhibits strong similarities with experimentally observed dissipation suggesting that drag is likely to be the main source of energy dissipation. Yet, experimental values don't match exactly with the predicted ones. A cause of this discrepancy could lie in the fact that drag creation is more complex than the description proposed by this model. Indeed, as studied by Poupardin \textit{et al.} (2012), in the case of rigid plates in waves vortices are mainly created at the edges \cite{Poupardin2012}. Such effect could arise in the case of the oscillating elastic plate explaining the discrepancy. The precise quantification of the importance of these phenomena is the next step in order to fully understand the behaviour of the interaction of surface water waves and fully-submerged elastic plates.

\section{Conclusion}

 This preliminary study shows that the chosen design of fully-submerged elastic plate predominantly transmits incoming energy while reflecting almost no energy. However, significant energy dissipation (up to 40\%) is found to occur from the wave/membrane interaction at larger wave frequencies for which the wavelength is less than two times the plate length.
\\ Relatively simple theoretical models are employed to investigate the possible causes of dissipation. Calibrating those models experimentally, it is found that internal damping is unlikely to play an appreciable role in the experiments carried out. In contrast, the viscous dissipation model presented Section~\ref{Section: ResultsC} shows remarkable similarity with the empirically observed dissipation presented Fig~\ref{Coeff_nrj_times2}: modelled and experimentally observed dissipation values have similar magnitude and exhibit a marked dissipation peak for wavelengths close to $\lambda=1.5L$.\\
\begin{figure}[!h]
\centering
\includegraphics[width=86mm]{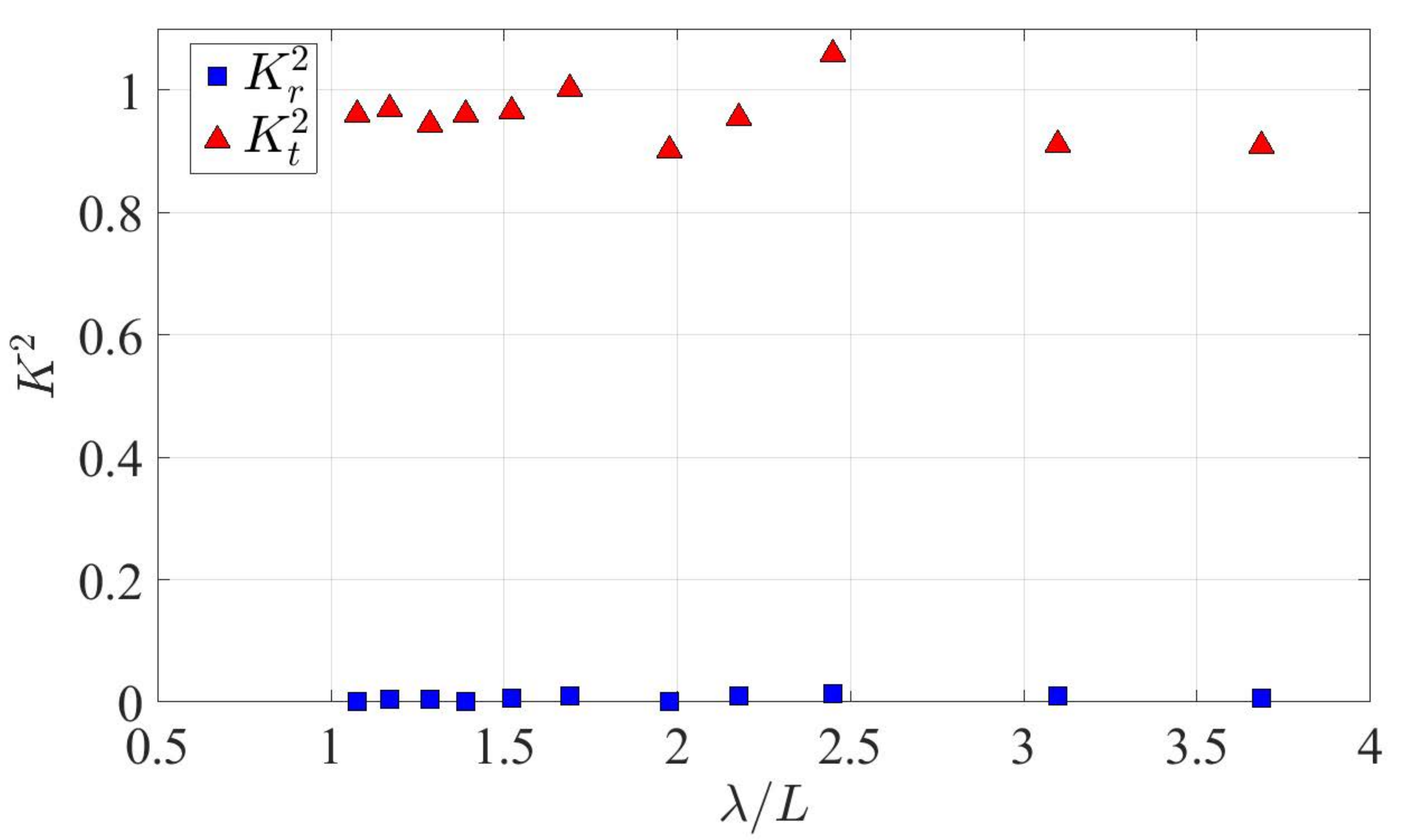}
\caption{Coefficients of reflection (blue squares) and transmission (red triangles) in energy as a function of the wavelength, $\lambda$ normalized by the length of the plate. The support poles don't interact much with the wave field.}
\label{K_potos}
\end{figure}
In conclusion, the proposed setup allows to highlight and quantify the variety of relevant physical behaviours in the wave-plate interaction. Nevertheless, with a view to wave energy absorption, it is clear that the membrane material and thickness do not allow the observation of internal energy dissipation (which would be representative of the system of wave energy harvesting capabilities). It is expected, however, that the simple dissipation models employed here will assist in the choice of more pertinent material and dimensions. Furthermore, this study concentrates on a 2 dimensional behaviour of waves but for wavelengths shorter than the tank width, diffraction and radiation by the elastic plate results in waves propagating in lateral direction. Using 3D Fourier transform could allow to quantify the part of energy propagating laterally to complete this study.

\appendices
\section{Influence of poles on waves}

To quantify the influence of the support poles on the wave field, reflection and transmission are measured for various wavelengths. As for experiments with the elastic plate, waves are imposed for 60 s to reach a stationary state before recording for 15 s. The duration of the acquisition is reduced compared to measurements with the elastic plate in order to simplify the treatment. Results are presented in Fig.~\ref{K_potos} and show that poles have few interaction with the wave field for wavelength superior to the length of the plate. 
The reflection observed corresponds to noise due to limits of the Schlieren method.
\begin{figure}[!h]
\centering
\includegraphics[width=86mm]{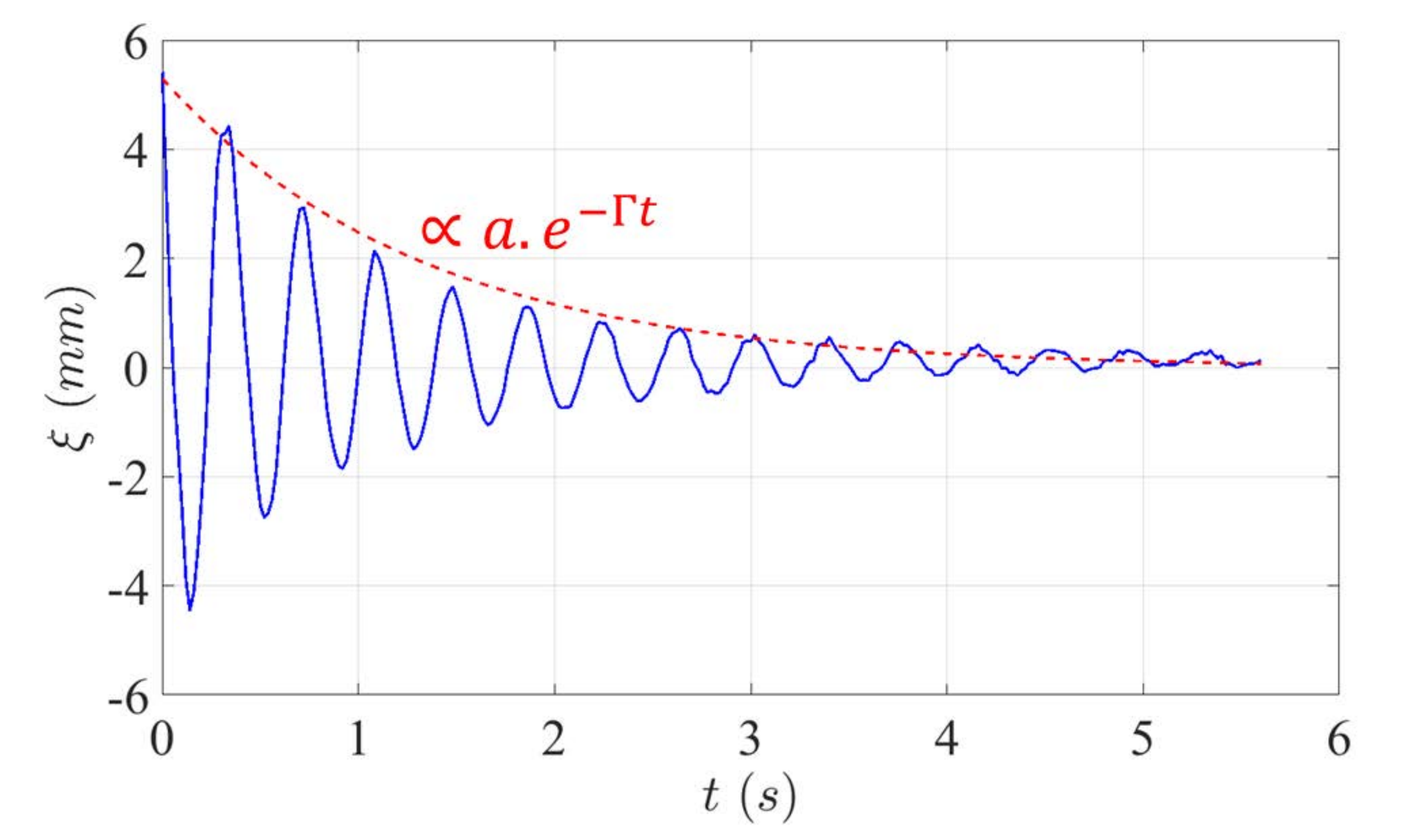}
\caption{Elevation of the edge of the elastic plate, $\xi$ as a function of time, $t$. The blue continuous line corresponds to the effective motion of the edge and the red dotted line corresponds to the exponential fit of its envelope. The inverse of the characteristic time of the fit, $\Gamma$, corresponds to the damping coefficient.} 
\label{expdecay}
\end{figure}

\section{Measurement of an internal damping coefficient using a free oscillation test}

In order to evaluate the influence of internal damping on wave energy dissipation, the internal damping coefficient, $\Gamma$, has to be measured. To do so a free oscillation test in air is performed. The thin elastic plate is attached horizontally and filmed from the side at 50 frames per seconds. Its free edge is lifted above its initial position of around 1 cm and released. The edge of the thin plate oscillates with an exponentially-decreasing amplitude. The characteristic time of damping allows to determine the damping coefficient $\Gamma$ as presented Fig.~\ref{expdecay}. For the sake of precision this experiment is performed 5 times and gives $\Gamma=$0.72$\pm$0.09s$^{-1}$.




\bibliographystyle{IEEEtran}

\end{document}